# Towards A Learning-Based Framework for Self-Driving Design of Networking Protocols

Hannaneh Barahouei Pasandi, Tamer Nadeem

Dept. of Computer Science, Virginia Commonwealth University, Richmond, VA 23284, USA

barahoueipash,tnadeem@vcu.edu

## ABSTRACT

Networking protocols are designed through long-time and hard-work human efforts. Machine Learning (ML)-based solutions have been developed for communication protocol design to avoid manual efforts to tune individual protocol parameters. While other proposed ML-based methods mainly focus on tuning individual protocol parameters (e.g., adjusting contention window), our main contribution is to propose a novel Deep Reinforcement Learning (DRL)-based framework to systematically design and evaluate networking protocols. We decouple a protocol into a set of parametric modules, each representing a main protocol functionality that is used as DRL input to better understand the generated protocols design optimization and analyze them in a systematic fashion. As a case study, we introduce and evaluate Deep-MAC a framework in which a MAC protocol is decoupled into a set of blocks across popular flavors of 802.11 WLANs (e.g., 802.11 b/a/g/n/ac). We are interested to see what blocks are selected by DeepMAC across different networking scenarios and whether DeepMAC is able to adapt to network dynamics.

## 1 INTRODUCTION

The proliferation of the current Internet and mobile communications networked devices, systems, and applications has contributed to increasingly large-scale, heterogeneous, dynamic, and systematically complex networks. A very recent forecast from International Data Corporation (IDC) estimates that there will be 41.6 billion connected wireless devices in 2025 [11]. This significant growth and penetration of various devices come along with a tremendous increase in the number of applications supporting various domains and services. The increasing availability and performance demands of these applications suggest that "general-purpose" protocol stacks are not always adequate and need to be replaced by application tailored protocols. Ideally, protocols (e.g., for routing, congestion control, video streaming, etc.) will perform well across the entire range of environments in which they might operate. Unfortunately, this is typically not the case; a protocol might fail to achieve good performance when network conditions deviate from assumptions implicitly or explicitly underlying its design, or due to specific implementation choices by domain experts. However, while this approach is increasingly becoming difficult to repeat, these designed protocols are deeply rooted in inflexible, cradle-to-grave designs, and thus unable to address the demands of different network characteristics and scenarios [5, 9]. We believe that MAC protocols have the potential to outperform their human-designed process in certain scenarios as shown in [4]. Protocols designed using ML are not limited to human intuitions and may be able to optimize control-plane traffic and channel access in yet unseen ways.

Unlike others, we focus on decoupling protocol features /functionalities in different technologies and scenarios. To accommodate our idea, we use control parameters. Taking physical layer as an example, no single physical-layer design can work well under all scenarios, hence the natural response of the standards bodies has been to specify designs with a large number of control parameters ranging from modulation order and coding rate, to OFDM sub-carrier spacing and cyclic prefix length, to transmit power, etc., such that a medium can be tuned to the specific deployment scenario in the field. Each of these parameters has numerous settings leading to a large number of choices, and it becomes extremely difficult for domain experts to design a control algorithm that chooses the right algorithm depending on the scenario and the varying network conditions. Therefore, due to heterogeneity and dynamically changing characteristics of networks (e.g., IoT), network protocol design requires a new approach in which control rule optimizations are not only based on closed-form analysis of isolated protocols, but are based on high-level policy objectives and a comprehensive view of the underlying components. Thus, it has now become crucial to re-engineer protocol designing process and shift toward a vision of an intelligent designing process that adapts and optimizes network protocols under various environment contexts such as device characteristics, application requirements, user objectives, and network conditions.

Given that networking field often deals with complex problems that require efficient solutions, exploiting Machine Learning (ML) techniques for solving these problems looks promising. The value that such techniques can bring to the protocol design process comes from the flexibility of the protocol design process to construct protocols that can learn from their past experiences, and the ability to respond to network dynamics and environment conditions in real-time. Among ML techniques, Reinforcement Learning (RL) is suitable for the unknown environments where decision-making ability is crucial. Recently, Deep learning (DL) and Deep Reinforcement Learning (DRL) [35] techniques have been applied to various protocol and radio optimization tasks including routing [24], congestion control [80], MAC [55] [82] and frequency estimation in PHY layer [90], just to name a few.

To the best of our knowledge, the current efforts in applying DL to enhance protocol performance focus only on *tuning* or *controlling* protocol parameters. Applying such DL-based techniques can reduce manual human-based efforts to tune protocol parameters. However, the black-box nature of DL-based techniques leave us little insight about how they work. Joseph et al. [32] show how to design a DL-based control algorithm to jointly control two parameters namely modulation order and transmit power scaling. In their work, they show that applying DL technique may work well to control the two aforementioned parameters, but depending on the context (different devices, throughput targets, etc.,) it becomes extremely complicated to get enough insights about how black-box DL technique works, even only for tuning two parameters from a large set of available control parameters. We believe that optimizing a protocol performance goes beyond individual protocol parameter *tuning*.

In this paper, we propose a novel DRL-based framework that is not only capable of tuning protocol parameters, but also optimizing the main functionalities for each protocol. In the proposed framework, a protocol is decoupled into a set of parametric modules as DRL inputs, each representing a main protocol functionality referred as *Building Blocks*(BBs). This modularization technique helps to better understand the generated protocols, optimize the protocol design, and analyze them in a systematic fashion. These BBs and a set of other parameters are fed into the DRL agent as the input. The DRL agent then is able to learn what protocol blocks (components) are important to be included or to be excluded from the protocol design. Note that our proposed framework is generic for designing networking and communication protocols in all layers of network stack. As a case study in this paper, we narrow down our focus to propose a DRL-based framework for designing wireless MAC protocols *hereafter* `DeepMAC` as a version of the proposed framework.

In `DeepMAC` framework, MAC protocols are decoupled into a set of parametric modules, each representing a main functionality across popular flavors of 802.11 WLANs (IEEE 802.11 b/a/g/n/ac amendments). As we showcase in Section 6, the DRL agent learns that when the load of the network is very low, it could eliminate control and sensing mechanisms (ACK and Carrier Sensing blocks, respectively) to increase the throughput of the channel by reducing the bandwidth overhead and waiting time introduced in these mechanisms. Therefore, this framework could serve as a tool for protocol designers to re-think the blocks used in a designed protocol. In addition, our framework could be utilized as a multi-variant optimization tool that helps in alleviating the current protocol design process. By using this framework, domain experts provide the required specifications (objective) for a specific scenario as DRL input and could identify/capture the role that each protocol component (block) plays in varying scenarios for different objectives. It could also help domain experts to get insights about the relation between different protocol components for different objectives, although such components may not have a direct dependency/relation to each other if considered alone.

**Contributions** The main contributions of this paper are summarized as follows.

i) We motivate a novel learning-based protocol design approach that not only tunes the protocol parameters but also optimizes the protocol design by leveraging the concept of demodulating a protocol into its set of main functionalities referred as building blocks.

ii) To show the feasibility of our framework, we propose `DeepMAC`, a novel deep reinforcement learning-based framework that targets the design of 802.11 MAC protocols based on the given networking scenario. Evaluation results shows that `DeepMAC` can intelligently selects the optimum protocol design for a given objective (higher throughput) under different scenarios and outperforms conventional conventional protocols e.g., CSMA/CA. By using the demodulating concept, we are able to interpret `DeepMAC` behavior under different scenarios.

## 2 BACKGROUND

Reinforcement learning is a machine learning technique where the agent interacts with a time-variant *environment* that can be modeled as a Markov Decision Process (MDP), a Partially Observable MDP (POMDP), a game, etc. The core components of RL technique are the **environment**, **reward** ($r$), **possible set of actions** ($\mathcal{A}$) and **states** ($\mathcal{S}$). The state is the agent's perception of the environment and is defined based on the agent's sensory information. The agent selects an *action* in a given *state* and receives a *reward*. Actions are the agent's methods which allow it to interact and change its environment, and thus transfer between states. The *policy* $\pi$ prescribes actions to take in a given state. We can then value a given state $s$ and a policy $\pi$ in terms of expected future rewards.

Every RL algorithm must follow policy $\pi$ in order to decide which actions to perform at each state. RL algorithms are mainly categorized into on-policy and off-policy reinforcement learning. Algorithms which concern about the policy which yielded past state-action decisions are referred to as on-policy algorithms, while those ignoring it are known as off-policy. In other words, there are two main methods to solve RL problems: calculating the value functions or Q-values of each state and choosing actions according to those, or directly compute a policy which defines the probabilities each action should be taken depending on the current state, and act according to it. A well known off-policy algorithm is Q-learning [76], as its update rule uses the action which will yield the highest Q-value, while the actual policy used might restrict that action or choose another. The on-policy variation of Q-Learning is known as SARSA, where the update rule uses the action chosen by the followed policy. For a more thorough review of reinforcement learning please see [43, 66]. There are many different variations and assumptions that change the methods in a RL problem; here we focus on Q-Learning.

Q-Learning (QL) is one of the most popular off-policy algorithms in RL. It is also regarded as temporal difference learning that learns an *action-value* function to find a Q-value for each state-action pair. QL agent learns its optimal policy by exploring and exploiting the environment. At each time instant $t$, the agent observes the current state $s_t$ and chooses a proper available action $a_t$ from this state to maximize the cumulative reward in time instant $t + 1$. More formally, the Q-value of $(s_t, a_t)$ from the policy $\pi$ which is denoted as $Q^\pi(s_t, a_t)$ is the sum of discounted reward received at time $t + 1$ when action $a_t$ is taken in state $s_t$, and it follows the optimal policy $\pi^*$, thereafter. The Q-values are updated using the following rule known as one of the Bellman equation forms:

$$Q(s_t, a_t) \leftarrow Q(s_t, a_t) + \alpha[r_{t+1} + \gamma \max_{a_{t+1}} Q(s_{t+1}, a_{t+1}) - Q(s_t, a_t)] \quad (1)$$

where $\alpha$ is the learning rate and $\gamma$ is the discount factor. Intuitively, the above equation adjusts the long-term or delayed rewards for a given state, action and future state $s_{t+1}$ by weighting the previous Q-value estimate, the reward received and the best possible long-term reward obtained in the future state. The Q-values estimate can be tuned for seeking any delayed reward desired. For instance, to seek short-term rewards exclusively, we can set $\gamma = 0$, while for estimating the history of all rewards it would be $\gamma = 1$.

The capability of traditional RL techniques is limited by the curse of dimensionality, and it will be inapplicable in large-scale systems [70]. To overcome this problem, Deep Reinforcement Learning (DRL) [48] has thus been proposed. We obtain DRL methods when deep neural networks are used to approximate reinforcement learning components including value function, policy, and model. By integrating DL into RL, DRL uses Deep Neural Nets (DNNs) to overcome the curse of dimensionality and hence is able to solve large-scale problems effectively. When using neural networks with QL, the QL's table is replaced by the neural network that is referred to as the approximator function, and denoted as Q(s,a; $\theta$), where $\theta$ represents the trainable weights of the network. In deep QL, the state is given as the input and the Q-value of all possible actions is generated as the output. In this case, the Bellman Equation is used as the cost function which is squared error of the predicted Q-value and the target Q-value as follows.

$$cost = [Q(s_t, a_t; \theta) - (r(s_t, a_t) - \gamma \arg\max_a Q(s_{t+1}, a_{t+1}; \theta'))]^2 \quad (2)$$

Multi-agent reinforcement learning (MARL) [22] is an emerging paradigm in RL, allowing to take advantage of having multiple learners analyzing the environment at the same time. MARL is the integration of multi-agent systems (MASs) with RL and it is suitable for distributed systems. Besides issues in RL like convergence, there are new issues like multiple equilibria and even fundamental issues like what is the question for multi-agent learning to solve, whether convergence to an equilibrium is an appropriate goal, etc. In MASs, each agent not only learns to operate independently but also cooperates with others to achieve the best joint reward. The direct strategy is to apply independent Q-learning in each agent and consider other agents as a part of the environment. However, this approach limits the number of agents because it is computationally expensive to train every agent in the system. There are different approaches to design a MARL framework, but they are beyond the scope of this paper.

## 3 RELATED WORK

Recent breakthroughs in ML techniques have drawn the network community's attention. By having the ability to interact with complicated environments and decision making, ML techniques provide promising solutions for higher network performance. More specifically, RL has been utilized in tuning and optimizing many network sub-fields [45] such as Wi-Fi [26, 33] in PHY [25, 49, 53], MAC [39, 55], and upper layers [6]. Today, networks are getting more complicated, and traditional modeling and optimization methods are becoming less effective and inefficient. DL can find the relationship hidden in the training data, and then use it to make accurate predictions even in an unknown situation. Without the need for modeling, the data-driven nature of DL is an appealing feature in system design for complex networks. In the following, we summarize some of the research efforts which use RL and DRL for improving communication protocols in different network stack layers.

RL has been utilized in tuning PHY and MAC parameters in wireless sensor networks. S-MAC [84] applies RL to adaptively tune the duty cycle in which nodes form a virtual cluster in order to provide a common schedule between neighboring nodes and a small SYNC packet is exchanged between the neighbors to ensure synchronized waking period to reduce control overhead. RL-MAC [40] is another mechanism that adapts each frame active time and duty cycle based on node's traffic. Each node in RL-MAC not only considers its own local state but also infer the state of other neighboring nodes in order to achieve near optimal MAC policy. The local observation of a node includes successful transmission and reception of packets during the active cycle, while the neighboring observation infers the failed transmissions to inform the receiver about the missed packets. In this model, the RL agent's state is the number of queued packets for transmission at the beginning of a frame and the action is the reserved active time to be tuned by the agent. The reward in RL-MAC is the number of successful transmitted and received packets during the reserved active time. The authors of [75] use deep Q-learning [47] solution for the dynamic multi-channel access problem in the wireless sensor network. In their model, each channel can have two states good or bad that represent the low and high interference, respectively. The objective in this work is to detect when the channel is in the good state for transmission based on the previous channel-transmission status (i.e., success, collision, idleness) in order to optimize the reward functions (e.g., throughput). However, this simple two-state channel representation cannot capture all the interference that may affect MAC behavior.

In recent years, DRL has been used for Dynamic Spectrum Access (DSA) and high performance has recently been achieved by subdividing the task into the smaller sub-problems of channel selection, admission control, and scheduling [8]. Naparstek et al. [50] consider the problem of dynamic spectrum access for network utility maximization in multichannel wireless networks. The authors propose a novel distributed dynamic spectrum access algorithm based on deep multi-agent reinforcement leaning. In their mechanism, the objective is to maximize a certain network utility in a distributed manner without online coordination or message exchanges between users/agents. The action of each agent at the beginning of each time slot is to select a channel and transmit a packet with a certain transmission probability. In [75], the DSA problem is formulated as a POMDP with unknown system dynamics. In this framework, a sensor at each time slot selects a channel to transmit data and receive a reward based on the success or failure of the transmission. The channel state is either in low interference, i.e., successful transmission, or in high interference, i.e., transmission failure. In particular, the action of the sensor is to select one of the M channels. The sensor receives a positive reward +1 if the selected channel is in low interference, and a negative reward −1 otherwise. The objective is to find an optimal policy that maximizes the sensor's expected accumulated discounted reward over time slots. The channel access problem in the energy harvesting-enabled IoT system is investigated in [15]. Consuming more power leads to poor sensor's performance due to its energy constraint. The prposed model consists of one Base Station (BS) and energy harvesting-based sensors. The BS as a controller allocates channels to the sensors. The BS's problem is to predict the sensors' battery states and select sensors for the channel access so as to maximize the total rate. Since the sensors are distributed randomly over a geographical area, the complete statistical knowledge of the system dynamics, e.g., the battery states and channel states, may not be available. Thus, the DQL uses a DQN consisting of two LSTM-based neural network layers. The first layer generates the predicted battery states of sensors, and the second layer uses the predicted states along with Channel State Information (CSI) to determine the channel access policy. The state space consists of channel access scheduling history, the history of predicted battery information, the history of the true battery information, and the current CSI of the sensors. The action space contains all sets of sensors to be selected for the channel access, and the reward is the difference between the total rate and the prediction error.

To investigate the multi-channel problem, a DRL-based adaptive modulation and coding scheme is developed in [89] for primary users to learn the interference pattern of secondary users. The authors in [13, 93] also investigated the multi-channel access problem where the users aim to maximize their own throughput by learning the channel characteristics and the transmission patterns of the "primary" users. LTE in unlicensed spectrum using licensed assisted access LTE (LTE-LAA) is an approach to overcome the wireless spectrum scarcity. Tan et al. [67] developed DRL-based algorithms to learn WiFi traffic demands by analyzing WiFi channel activity, e.g., the idleness/business of WiFi channels, which can be observed by the LTE system via monitoring WiFi channels. Based on the learned knowledge, the LTE system can adaptively optimize LTE transmission time to maximize its own throughput and meanwhile to provide sufficient protection to the WiFi system.

Similarly, RL has been utilized to optimize packet transmission rate of MAC protocols in wireless networks. ALOHA-Q [16] combines slotted ALOHA and Q-Leaning. In this work, time is divided into equal time slots that transmissions are only allowed at the beginning of slots, and each packet is assumed to be the length of multiple time slots. In ALOHA-Q, each node stores a Q-value for individual time slots in which the agent's action is to select the slot with highest

Table 1: Examples of RL Formulation For Networking Protocols at Different Layers

| Network Layer | Function / Sub-Layer | Input (State) | Objective (Reward) | Description (Action) | RL Model Used | Related Work |
|---|---|---|---|---|---|---|
| Data Link | Dynamic Spectrum Access | • The state of the channel whether transmission is successful or not. | • Higher sending rate. | • Choosing to transmit a packet on specific channel. | RL, DRL | [12, 50, 75] |
| Data Link | Customized MAC | • Set of core MAC functionalities referred as building blocks. | • Maximizing the throughput. | • Selecting among set of MAC protocol building blocks. | RL | [55–58] |
| Data Link | Contention Window | • Transmit or wait to send a packet.<br>• Channel observation: idle, busy. | • Maximizing throughput.<br>• Allocation fairness. | • Transmitting a packet or waiting. | RL, DRL | [3, 47] [74] |
| Network | Routing | • Successful/unsuccessful packet transmission from a node to its neighbors. | • Numerical positive values for successful transmissions.<br>• Negative values for link costs for unsuccessful transmissions. | • Forwarding a packet from a node to a neighbor node.<br>• Broadcasting a packet by a node to discover new neighbor nodes. | MARL, POMDP | [21, 52] |
| Transport | Congestion Control | • Latency.<br>• Buffer occupancy level.<br>• Active flow. | • Maximizing throughput while penalizing loss and latency.<br>• Fair bandwidth allocation for each host while minimizing switch queues. | • Adjusting sending rate. | RL, DRL | [19, 63] |

Q-value for its next transmission. The reward function is a simple numerical value of +1 for successful transmission and −1 otherwise, which is used to adjust Q-values after each transmission. Similarly, the work in [7] uses Q-learning to enable wireless nodes to learn their own optimal transmission strategy using the historical sensory information that defines the node state including the number of transmission in the previous slot, number of consecutive idle or useless slots, and number of consecutive collisions. The action of each agent at each slot is either to transmit a packet with a certain probability (i.e., persistent probability) or to postpone it to the next slot. The reward function is a numerical value that is equal to 1 in case of successful transmission, negative constant $C$ in case of a collision, or 0 for no transmission. Recently, Alfredo et al. [17] proposed a policy-baced RL approach to improve slotted ALOHA in terms of fairness. Unlike Aloha-Q, they consider a dynamic time frame, and each node learns the best time slot to transmit a packet based on its local policy tree.

There are prior works that exploit DNNs technique to optimize MAC protocol design using supervised learning [34, 38, 59, 61, 62, 78, 79, 81, 83] and unsupervised learning techniques [10, 14, 51, 92]. [55] [86]. There are few prior works that focus on MAC protocol design optimization in wireless networks [55, 57, 87]. Authors in [86] investigate a DRL-based MAC protocol for heterogeneous wireless networking. Their method objective includes maximizing the sum throughput and maximizing $\alpha$-fairness among all networks. In [3], the conventional CSMA/CA protocol is improved for densely deployed WiFi networks, in which DRL is adopted to learn the optimal CW for each WiFi node to improve overall throughput. Authors in [55, 57] target IEEE 802.11 MAC protocol design. They decouple a MC protocol into it's set of building blocks. The agent then selects the appropriate protocol blocks based on varying network condition. Valcarce et.al [72] investigate the problem of joint MAC signaling and channel access using MARL technique.

RL has been utilized for optimizing routing protocols [23, 68, 71, 85]. Routing management is a crucial aspect for traffic control as the poorly chosen paths can lead to network congestion, and then the following retransmissions of the lost packets may further aggravate the congestion. In traditional routing protocols, the main concept is to choose the path having the maximum or minimum value or metric e.g., the Shortest Path (SP) algorithm. The traditional SP algorithm has the problem of slow convergence, which is not suitable for dynamic networks since the slow response to the network changes can lead to severe congestion. Authors in [20] propose SAMPLE a collaborative reinforcement learning approach to enable groups of reinforcement learning agents to solve system optimization problems online in dynamic, decentralized MANETs. In SAMPLE, agents have both internal (non-visible to neighboring agents) and external states. The state in SAMPLE is monitoring the combination of attempted unicast transmissions, successful unicast transmissions, received unicast transmissions, received broadcast transmissions, received (overheard) unicast transmissions. In the proposed protocol, actions can be performed locally at a node, or among nodes as a collaborative system. An action in SAMPLE is to broadcast a packet by a node to discover new neighbor nodes or the selection of a next-hop neighbor to transmit a packet. There are two types of objective functions in the proposed approach. One of the objective functions gives a positive numerical value for successful transmissions while it penalizes link cost of unsuccessful transmissions with a negative value. The second objective function represents the cost of executing actions locally at a node. RL is also used for designing routing protocols in WSNs. For a comprehensive survey on RL for routing, interested reader is referred to [44]. Authors in [28] propose QELAR a RL-based routing protocol for routing underwater sensor networks. In QELAR the definition of system states is related to individual packets i.e., if a node has a packet the system state depends on that packet. The action is to forward the packet, and the reward function is based on the fact that each packet forwarding attempt consumes energy, occupies channel bandwidth, and contributes to the number of hops to the destination (i.e., delay). With this constant punishment, the agent is compelled to choose the relatively shorter paths to the destination, and hence, the routing delay is minimized.

There have been several efforts to apply RL to congestion control in specialized domains. Prior work in [65] employs RL to create a cooperative congestion control controller for multimedia networks. RL has also been used to solve congestion problems in wireless sensor networks [88]. Authors in [2] explore designing a TCP-style congestion control algorithm using Q-learning. The recent effort [31] proposes a DRL-based adaptive framework for congestion control. In this model, state is bounded histories of network statistics (e.g., sending rate, latency), action is periodically tuning the sending rates, and the reward is a linear function that rewards throughput while penalizing loss and latency. A promising approach to multipath congestion control is to use reinforcement learning. Deep Reinforcement Learning Congestion Control (DRL-CC) [80] algorithm jointly sets the congestion window for all active flows and all paths and achieves high fairness in a wired network scenario with multiple active flows. Authors in [77] propose and evaluate Remy tool that generates congestion control algorithms to run at the endpoints rather than manually formulate each endpoint's reaction to congestion signals. Remy is a heuristic search algorithm that maintains rule tables in which states are mapped to actions. Unlike RL-based approaches, Remy is static meaning if the actual network conditions changes, performance could potentially degrade substantially.

Although the aforementioned mechanisms differ in the details, their common objective is to optimize a protocol by *tuning and/or controlling* the protocol parameters. Our proposed approach is different from these works in the following respects. First, we argue that designing methods to boost protocol performance is not only about parameter tuning, but also to decide what functionality to include or exclude from the design. The novelty of our approach resides in the way our framework constructs a protocol from a set of building blocks. By decomposing the protocol into the set of mechanistic building blocks, we aim to better understand the design, the interdependencies among different protocol building blocks, and to ease the analysis of the protocols. In theory, a self-managed protocol design framework would identify where and when to execute design tasks. This process includes reconfiguring the protocol by choosing the *right* set of building blocks when necessary, predicting performance problems due to unpredictable network changes. In addition, our framework supports multi-variant communication objectives that can be *explicitly* defined by domain experts. By using this framework, domain experts provide the required specifications for a specific scenario as DRL input and could identify/capture the role that each protocol block plays in varying scenarios for different objectives. It could also help to get insights about the relation between different protocol components for different objectives even when such components may not have a direct relation to each other if considered alone.

In Table 1 we summarize additional works similar to what we have discussed about RL formulation for protocol enhancement in different network stack layers.

## 4 FRAMEWORK FOR NETWORKING PROTOCOLS DESIGN

### 4.1 Framework Overview

In this subsection, we describe our proposed reinforcement learning-based framework that can be generalized to design different types of protocols with different level of complexity for any layer in the networking stack. Figure 1 illustrates the proposed framework overview.

As shown, the RL agent selects a set of building blocks from the existing blocks, along with the network configuration, and the reward feedback as input, and designs a protocol accordingly. The designed protocol is then evaluated (e.g., through simulation), and a reward (e.g., mean throughput of the link) is calculated and sent back as a feedback that signals the agent about the current protocol performance. This process iterates until a protocol with optimum set of building for the current network/environment configuration is designed.

In the following subsections, we explain the main components of the proposed framework.

### 4.2 Protocol Building Blocks

A network protocol is structured into several layers. Each layer is broken into a set of blocks with its own specific functionality. Building blocks are a set of separated parametric modular components, each of which is in charge of one (or several) specific well-defined functionality [18, 46, 69]. The combination of different blocks and the interactions between them determine the overall behavior of a network protocol for a given environment. Once blocks and their interactions are established, network protocol could be represented as a graph, where the parameterized blocks are the vertices and the edges connecting the blocks represent the transition between them. By conducting the operations of individual blocks in an appropriate order, we are able to implement the protocol mechanisms.

Yet among the main challenges to develop flexible and reusable set of building blocks is how to decide on the level of granularity of each block, and to evaluate different block granularity levels of the same function. One approach could be a brute-force in which the interactions, relations, dependencies, and conflicts between every couple of blocks is defined by design experts. However, this approach is complex and time-consuming. Therefore, to address these issues, there should be further research on exploring how to reduce this complexity through approximation techniques, as well as whether this process could be automated.

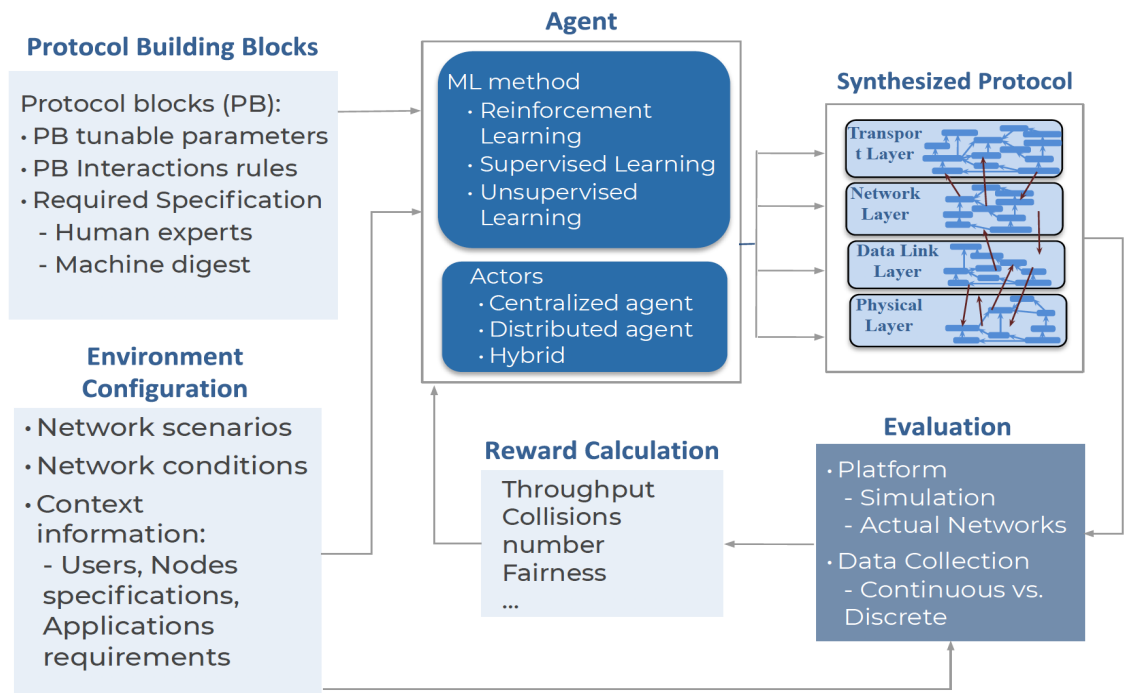

**Figure 1: The proposed RL-based framework**

Another main challenge is to check the sanity and correct execution order of the selected blocks that will be used in protocol design. Such sanity check requires to understand network condition effect on the selection of the blocks as well as logical dependencies among the blocks themselves to design a proper protocol. Therefore, we need to capture both the effects when formulating the building blocks which we discuss in more details in the following.

The network protocol is operated under a variety of conditions and environments, which trigger events causing the protocol to act. Following the modular design principle in context of protocol design, two main branches exist on how to divide protocols and define the components: Finite State Machine (FSM) and Data Flow. FSMs are graphical formalism that have become widely used in specifications of embedded and reactive systems. Their main drawback is that even for a moderate complicated system, they result in large diagrams. To overcome this, in the literature, prior works [69, 73, 91] use extended FSMs to model MAC protocols based on the block(modular)-based concept to add flexibility to protocol design, and reusability of the components. The existing (block-based) frameworks consider either a static way of connecting the components together, or add a limited degree of dynamics in binding the components together. Therefore, these models do not fully capture the interactions, relations, dependencies, and conflicts between different components.

In order to support the dynamic design of the protocols and the flexibility in selecting the optimum set of building blocks (components) by the RL agent in designing efficient protocols, capturing the interactions and dependencies between building blocks is of crucial importance. Therefore, when describing a building block, we should also capture the dynamic behavior of a protocol caused by different events. Building blocks should react to incoming events, conducting their main operation while interacting with each other in proper order. To check the sanity and correct execution of the blocks, one approach that we have used is to design a ***logic controller*** module in which the predefined rules are embedded using if-then-else statements. The dynamic behavior of a BB could be estimated if the input events are known since the behavior of the BBs is deterministic. To be exact, we could describe a building block and its dynamic behavior not only based on its main functionality, but also by capturing events and dependencies among blocks as the following tuple:

$$Block :< E, P, S, F, D > \qquad (3)$$

where E is the **E**vent is set of signals either provided by the hardware interrupt block, or coming from the upper layers, which triggers the block. P is a set of **P**arameters inside the block that could be tuned, S is the **S**tate of the block, F is the main **F**unction that is executed in the block, and D represents the possible **D**ependencies between a block with other blocks. In our approach, the dependencies are uni-directional meaning if *Block A* depends on *Block B* it only shows restrictions of A to B but not B to A. Blocks can have different types of dependencies between them that we have mainly categorized as: *strong*, *weak*, and *conditional* dependency. In our framework, a strong dependency is between those blocks that are tightly wired together and must be selected together in order to deliver their functionality properly. A weak dependency is identified, when a block can be updated/executed based on the output of the other block. However, if the latter block does not exist in the protocol design, the former block may use a predefined specification for its execution. Conditional dependency captures the logical order of the execution of the blocks. For example, if we have a block called Sending Packet to transmit a packet then it can be executed as it is, but if we have Carrier Sensing block present, then Sending Packet block is executed after Carrier Sensing.

To show an example using Tuple 3, lets consider Backoff mechanism as a single building block. A tuple that describes this block could be:

$$< ACK\_timeout, CW, Freeze/Countdown, BEB, ACK >$$

Acknowledgement time out referred as ACK_timeout in the above tuple is the event that triggers Backoff block, since it indicates that the frame transmission was unsuccessful. Contention Window (CW) is the parameter that can be tuned to different values to set the Backoff counter range, the state is the state of the Backoff counter which could either be counting down when the medium is sensed to be idle, or it could be frozen when the channel is sensed busy. The last component of the tuple is dependency which indicates whether Backoff mechanism is dependent on any other block. In this example, Backoff mechanism is strongly dependent on the ACK block (Backoff → ACK), since if there is no ACK block present, then there is no ACK_timeout event to trigger Backoff. Although the other direction does not hold, meaning we can use ACK without having a Backoff mechanism in a designed protocol.

### 4.3 RL Agent

The agent takes both the protocol building blocks and the network/environment configuration as inputs in order to utilize the building blocks and their interactions in designing the optimum protocol that fits with the network requirements using reinforcement learning mechanisms.

In RL, the agent training goes on by each Q-value function update. The offline RL training happens when the agent runs in variety of scenario and learned Q-values are stored such that the agent uses them when making new decisions. In online RL, typically, the agent starts from an initial state, and its training evolves over time meaning the agent does not have any prior knowledge (Q-value) when it starts from the initial state.

RL agent can be implemented in centralized or distributed approaches. Centralized agent means there is a single agent responsible for managing the protocol design task, and then the designed protocol is enforced to be used by all the nodes in the network. Decentralized approach assumes that multiple agents perform the task of learning based on their own knowledge, including what actions to take based on the current state and expectation of other agents' actions. Although in a distributed approach (i.e., multi-agent environment) each agent has the flexibility to design its own optimum protocol based on its characteristics and application requirements, instability throughout the network could happen as some agents may take random actions that can affect the learning process of other agents. On the other hand, while a centralized approach is simple and easy to control and manage, it becomes computationally expensive when number of nodes in network grows or the state space becomes large. Moreover, a centralized approach is not suitable for heterogeneous environment where different nodes have different objectives.

Combining the benefits of both centralized and distributed approaches, a hybrid approach could be designed for RL agents [27]. The work by Kraemer and Banerjee [36] proposes a hybrid solution in the context of Decentralized partially observable Markov decision processes (Dec-POMDPs) for modeling multi-agent planning and decision-making under uncertainty. More specifically, they proposed a centralized approach, where a group of agents can be guided at the same time by using a centralized algorithm via an open communication channel. Then, after the training, agents are allowed to communicate over a channel, and thus can operate freely in a decentralized manner. A similar approach could be adopted for a hybrid solution in protocol design context. A design strategy could classify the building blocks and their interactions into global and local blocks. Global blocks are the ones mainly responsible for designing the main skeleton of the protocol in which all nodes have to adapt/use the same selected blocks. On the other hand, local blocks are the ones that could be selected, adapted, tuned and configured individually based on the local context of each node.

Another design challenge especially for distributed and hybrid approaches is the communication strategy between agents in order to improve collectively their performance. The communication mechanisms between multiple agents could be mainly categorized to two main approaches; individual peer-to-peer channels and all-to-all channels. Peer-to-peer channels will enable peer agents experience similar conditions and targeting similar rewards to exchange their information with each other in order to speed-up the learning process and be able to converge to the optimum protocol within time constraints. Additionally, if the design includes a Long Short Term Memory (LSTM) that store previous experiences and observations of agents over time, an agent that is currently learning a new protocol may utilize information from another agent who dealt with similar task in the past. All-to-all channels will allow to send crucial information that will affect all agents in the system in order to adapt the protocol design such as changes in time constraints or discovery of new patterns in data.

### 4.4 Reward Function

The essential goal of RL agent is trying to learn a policy to select actions that maximize its expected rewards for state-action pairs. The agent learns that policy by interacting with the environment and observing the reward function in every state. Reward function varies from a simple performance metric such as total network throughput to complex formula. Using different reward functions will generate different protocol designs for the same network scenario. Therefore, reward function should be designed carefully based on the device characteristics and application requirements of the network. For example, with embedded devices such as Internet-of-Things (IoT) devices, power consumption would be more important than throughput and hence a protocol design that minimizes the retransmission is needed. In this case, the reward function needs to consider number of collisions and packet corruption in its formula in order to minimize the retransmissions.

Another important design decision is how to design and optimize the reward function which can be optimized globally or locally. In a global optimization, both centralized and distributed agents work towards optimizing the same goal, while in local optimization, distributed nodes can optimize their own goals. Each of these approaches have their own challenges. Different applications have different performance requirements. Therefore, defining the "right" global optimization objective is not straightforward. Optimizing the objective function relies on the assumptions that all end-hosts employ the same prescribed protocol. Thus, there is a limited support for network heterogeneity, as well as, fulfilling different applications' objectives. On the other hand, each node in a distributed optimization tries to optimize its own objective function in which it might not converge. As an example, if each node wants to maximize its own throughput, it sends as much packet as possible without having any idea how happy or unhappy the other nodes are. In [64], authors bring a a symmetric parking-lot topology example as a classical congestion control problem. They use PCC [19] to generate congestion control protocols in which each flow explicitly performs local optimization of an objective function. In this example, authors show that the link allocation found by PCC can vary widely, depending on the initial rates or which flow starts first, even though they all last forever once they get started which leads to unfairness of link allocation. Therefore, it is very important to understand which reward optimization works better depending on the given task.

## 5 DEEPMAC: A CASE STUDY FOR 802.11-BASED MAC PROTOCOL DESIGN

MAC protocols need to be designed with a rich set of requirements in order to satisfy the needs of the overlaying applications and network conditions. The huge diversity of possible network conditions implies that even a protocol that works well across a wide variety of network conditions may suffer from bad performance on other networks. In case of MAC protocols, due to the limited channel resources and a large number of devices accessing the channel, it is desirable that the MAC protocol minimizes the time wasted due to collisions or exchange of control messages. In addition, it is required that the effective throughput remain high irrespective of the traffic levels. Overwhelmingly, the main challenge is the dynamicity of network conditions (e.g., nodes entering and leaving). Thus, it is imperative that the MAC protocol can be easily scalable and adjusted delicately to the changing environment with little or no control information exchange.

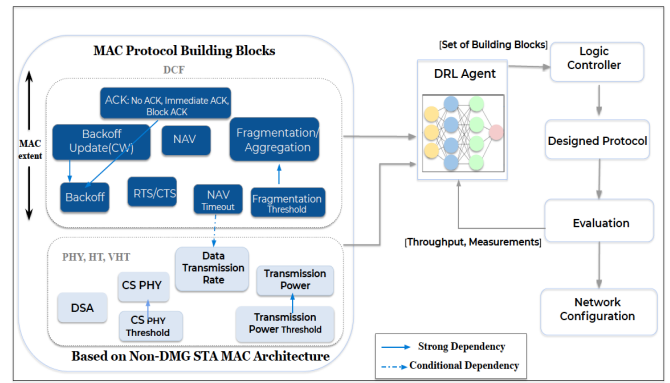

Figure 2: `DeepMAC` framework

To overcome the aforementioned challenges, we implement and evaluate `DeepMAC` that is a DRL-based framework that optimizes MAC protocols for a given networking scenario. In Q-learning, if the combinations of states and actions are too large, the memory and the computation requirement for Q will be too high. Therefore, We use a deep Q network (DQN) to approximate the Q-value function. Figure 2 shows `DeepMAC` framework and its key modules that aim to optimize the design of wireless MAC protocols. We describe the key modules in the following subsections.

## 5.1 DeepMAC Building Blocks

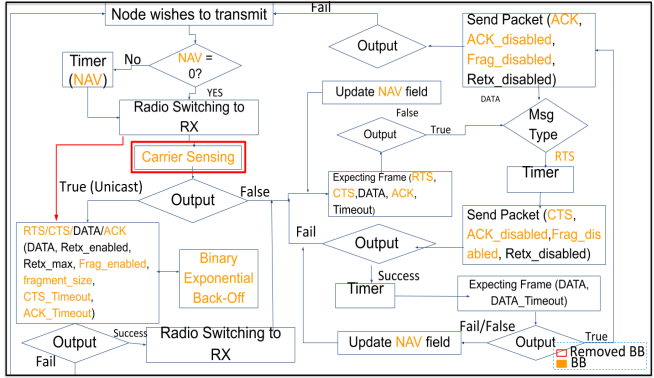

**Figure 3: Realizing IEEE 802.11 DCF using coarse-grained MAC blocks and their corresponding parameters**

In our framework, we have extracted a set of MAC protocol blocks from Wireless LAN Medium Access Control (MAC) and Physical Layer (PHY) Specifications [29] which includes the basic MAC DCF functionalities across all 802.11a/b/g/n/ac amendments that are shown in Figure 2. The realization of IEEE 802.11 DCF using these coarse-grained MAC blocks (namely Carrier Sensing (CS), ACK, Backoff, RTS/CTS, Fragmentation, Packet Transmission) and their corresponding parameters is shown in Figure 3. The modular concept of building blocks makes protocol design more flexible. As shown in the figure, if the RL agent does not select Carrier Sensing block in the protocol design for a given scenario, this block can be removed and the protocol execution sequence switches to the next logically selected block. As mentioned earlier and described later in more details, in our framework the logic controller checks the sanity and execution order of the blocks using embedded if-then-else rules. Having established a number of potential building blocks, DRL agent takes these blocks and a history of the average channel throughput as its main inputs as described later.

## 5.2 DeepMAC As A Reinforcement Learning Problem

`DeepMAC` uses Reinforcement Learning (RL) along with a deep architecture to learn the best set of protocol blocks for different scenarios. In `DeepMAC`, we consider a *centralized* learning agent for the design of 802.11 MAC protocols. This centralized agent, in practice, can be placed on a single supernode (e.g., the Access Point) that periodically updates its model. The supernode decides the selected set of MAC layer blocks and parameters to be used by all the other nodes in the network.

**Reward function** The designers need to specify the communication objective according to corresponding use case, device category, etc., under different scenarios. In the DRL framework, this defines the reward function. For example, for a battery-constrained IoT device hat cares for maximizing the throughput while minimizing spent energy as discussed in [37, 60], a designer may specify to optimize the following reward function:

$$w_0 \text{ (number of successful transmitted bits)} - w_1 \text{ (energy spent per bit)}$$

where $w_0$ and $w_1$ control relative tradeoff between these two conflicting components. In `DeepMAC`, the reward function is the *average throughput* of the link. Although such reward objectives can change based on the provided scenario by the protocol designer.

**State, Action** The *state* of the agent is a vector of numerical representation of the set of the building blocks, and a history with a fixed link of the average link throughput values which are used as part of the input for `DeepMAC` agent. In this set, for each block, a value except 0 indicates that the corresponding block is included in the protocol design (each of the elements in the input vector can have different values which indicate what parameter or algorithm/method/mechanism should be used in the design), while 0 means the block is completely excluded from the design. The *action* in this framework is the act of choosing the next state among all the available states from the current state such that the *reward* is maximized. Given the input, the output of the simulator is the average throughput of the channel which is considered as the reward of the DRL agent for the selected building blocks at the current step. We assume that all nodes employ the same prescribed protocol using the selected blocks by the centralized agent. The agent takes both the protocol design blocks and history of the reward as inputs, and outputs the best combination of building blocks for the current scenario that maximizes the reward.

**DRL agent architecture** The neural network we adopt is equipped with three hidden layers and an output layer. We find through our experiments that this simple architecture can yield satisfactory performance, and increasing the complexity of the neural network does not contribute to performance improvements while inducing more training overload. The data is flattened before going through the hidden layers which utilize Relu as the activation function. The output layer consists of multiple neurons, each producing the Q-value of the corresponding action.

## 5.3 Logic Controller

In network protocols, some functional blocks are dependent on each other. In our framework, the logic controller is responsible to check the sanity of a generated protocol. More specifically, the designed logic controller checks a) the blocks' execution sequences b) their interdependencies, and c) interaction rules among blocks to ensure logically correct protocol design. We extracted the interdependencies among different blocks from PHY and MAC specification [29], and incorporated them in the logic controller using if-then-else rules. We provide the following examples to describe some of such dependencies in the following. As shown in Figure 2, NAV (Network Allocator Vector which is considered as virtual CS mechanism) is conditionally dependant on RTS/CTS block. There are two methods to set the NAV parameter: a) by reservation information distributed through the RTS/CTS Duration field and b) by information provided in the Duration/ID field in individually addressed frames. Therefore, in our framework, NAV is set based on the RTC/CTS block if this block is available (selected by the agent). Otherwise, it is set based on the latter approach.

## 5.4 Network Configuration and Designed Protocol Evaluation

Integrated with protocol design element would be the network scenarios and conditions, such as communication medium types and node mobility. Different scenarios have different assumptions and requirements that need to be captured when designing a protocol. To evaluate DeepMAC framework, we developed an event-driven simulator using C++, while having the ns-3 design in mind. Our simulator mimics the MAC protocol of ns-3, but it is flexible to support the decomposed building blocks[1], and consequently the design of MAC protocols. Each building block is considered as a module and the agent decides about the inclusion and exclusion of the block as a part of protocol design. As input, the simulator takes the values of building blocks from the DRL agent that passed the logic controller check of finding any type of conflict or interdependency between them. It also receives the network configuration parameters including the number of nodes, level of noise, etc.

## 6 DEEPMAC EVALUATION

This section presents the numerical results and evaluation of DeepMAC regarding a) *convergence* b) *average throughput* enhancement and c) *block selection* by the agent under different scenarios, respectively. We assume that the supernode (centralized agent) uses hardware accelerators which can reduce the execution time by an order of magnitude and comfortably meet the real-time requirement. We clarify that our RL agent uses an online approach where it has no prior knowledge about the underlying environment.

### 6.1 Simulation Configuration

We consider an ad hoc network where individual nodes communicate with each other directly. To carry out our simulations, we use our event-driven simulator. Table 2 summarizes the simulation configuration parameters used in our experiments. The nodes are static and are randomly scattered in a 200x200m area. In our experiments, we consider eight different networking scenarios described in Table 3. The low load corresponds to an under-saturated network with 5 nodes, and average packet generation rate of 8 packets per second, while scenarios with high load represent close to saturated, and saturated networks with 20 to 50 nodes, and average packet generation rate of 470 packets per second. With regards to noise, when noise is not present, the received packets are assumed to be delivered with no error with probability 1, while when noise is present a fixed bit error rate (BER) of 0.0001 is considered. Scenario 1, for example, corresponds to a network having 5 nodes with a low traffic load that represents an under-saturated network while noise is absent. Table 4 includes the blocks and their associated algorithm, mechanism, or parameters and the default values that are used by DeepMAC framework for the experiments. Some blocks have different algorithms or parameters. As an example, if the fragmentation block is not selected by the agent, then the frame size remains 1500 bytes in the corresponding scenario. Otherwise if selected, the frame size varies. In order to see what blocks are selected by the agent in different scenarios, the evaluation for each scenario is performed 20 times. We then collected those blocks that are selected together more frequently than others by the agent over 20 rounds of repeating each scenario.

### 6.2 Learning Convergence Rate

We first evaluate the learning convergence rate of the RL agent to the goal state. We measure the convergence rate based on the number of the times that the agent selects the goal state which includes those building blocks that contribute to the highest throughput of the channel for a given scenario. Although, the goal is unknown to the agent, as a ground truth we assume that we know the goal state in order to count how many times the agent visits this goal state in every episode. Each episode contains 100 steps, and each step is considered as the action of selecting a state and updating the corresponding Q-value.

In addition, network channel condition is dynamic which makes the previous optimal policy learned by the agent inefficient. Therefore, the agent must find a new optimum policy to adapt to new changes. As it is not efficient to retrain the RL agent each time the characteristics of the channel change, we want to investigate whether the previously learned Q-values are helpful or harmful for converging to the new optimum policy. We evaluate RL convergence under eight different scenarios before and after the network in trained for that particular scenario. We carried out a set of experiments in which we compare the convergence of the agent for all combinations of different scenarios. Due to the space limitation, we only show how to analyze the results of Scenario #1 and #5, while the results of other scenarios resemble similar behavior. To evaluate the convergence rate when the agent starts to learn from an initial state, we let the agent learn the optimum goal states in Scenario #1 and #5, respectively. We then initialize agent Q-Values with those values learned from a different scenario namely Scenario #4 to evaluate whether the learned Q-values from Scenario #4 helps the agent to converge faster or slows down the convergence compared to the case when all Q-values are set to zero.

---
[1]High-end simulation tools (such as Opnet, NS-3, etc.) have the ability of reproducing with an accuracy of implementation. However, such tools do not support our building block decomposition concept properly.

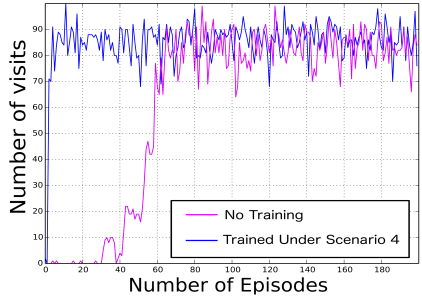
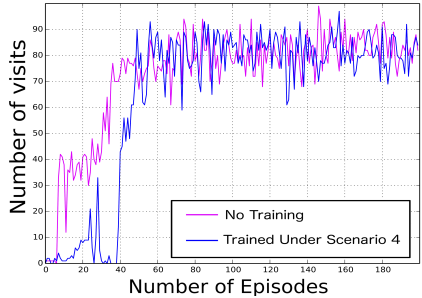
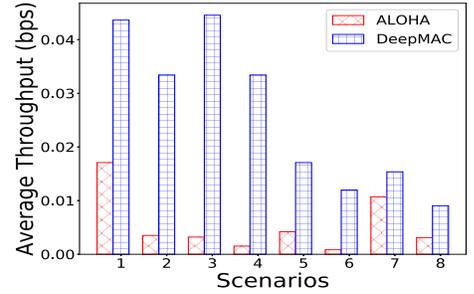

Figure 4: Number of goal state visits per episode for Scenario 1.

Figure 5: Number of goal state visits per episode for Scenario 5.

Figure 6: Throughput Comparison of **DeepMAC** against ALOHA.

Figures 4 and 5 show the convergence of the RL agent in Scenario #1 and #5 without any prior learned Q-values and with Q-values learned from Scenario #4, respectively. As shown in Figure 4, we can observe that the previously learned Q-values can actually help the agent for faster convergence to the goal state, while the agent takes more steps to converge when it is not trained (all Q-values are set to 0 initially). In the second scenario shown in Figure 5 the previously learned Q-values from Scenario #4 are used to train the agent in Scenario #5. As we observe, using the previously trained values will delay the convergence of the agent when compared to a scenario without training. This is because the reward values of these two scenarios have a significant difference meaning the agent needs additional number of steps to get out of the sub-optimal states and find the new goal state. Therefore, when there is a drift in network condition we first should be able to measure the significance of the drift and decide whether the agent should start from an initial state or it is helpful to making decisions based on the previous observations. However, identifying significant network drifts requires to be carefully investigated for wide range of scenarios that we plan to address this in our future work. In addition, as shown in the graphs, the RL agent converges approximately after 50 episodes in both the scenarios. The RL agent number of visits per episode fluctuates after convergence and never goes all the way up to 100 visits (which is the maximum number of visits in an episode). These oscillations are caused by the exploration versus exploitation ($\epsilon$-greedy) action selection policy. This behavior is expected since Q-learning with $\epsilon$-greedy approach will converge to the Bellman equation, but its *behavior* will not converge due to random action selection with a frequency of $\epsilon$ ($\epsilon = 0.05$). In the literature, one approach to smooth the curve in the graph is to decrease $\epsilon$ value over time. However, in our approach we have selected a fixed $\epsilon$ value experimentally.

### 6.3 DeepMAC versus ALOHA

In ALOHA, a node with a packet to send simply transmits. However, the simplicity of Pure ALOHA (referred as ALOHA hereafter) comes at the price of poor performance, with a maximum throughput of only 18% of the available bandwidth. As a result, several variants of ALOHA have evolved over the years to allow more efficient sharing of common channels in untethered networks. We are interested to see whether DeepMAC framework is able to design optimum simple ALOHA-like protocols for a given scenario. We compare the performance of ALOHA under eight different scenarios with DeepMAC. To design ALOHA-like protocols, DeepMAC is set to use a limited set of building blocks (namely ACK, Backoff, Contention Window, Carrier Sensing, Data Transmission Rate) that could only construct an ALOHA protocol. ALOHA is originally designed for wired networks and does not include ACK. However, we have modified ALOHA to fit the wireless network configuration. The intuition behind this comparison is to see whether the agent is able to perform better when it has the same or even less number of functionalities (building blocks) compared to ALOHA. To this end, Figure 6 illustrates higher throughput gains of DeepMAC against ALOHA in all the scenarios. Due to its self-adaptive characteristics, DeepMAC uses the set of blocks in different scenarios to enhance the throughput. For example, in Scenario #1 where the network load is low, `DeepMAC` removes ACK and Backoff components to gain a higher throughput, while in Scenario #6 where the traffic load is high and noise is present it includes ACK and Backoff to enhance the number of successful transmissions. Therefore, this consideration shows that having an intelligent agent that can decide in what scenarios which component can be beneficial is important.

**Table 2: Simulation configuration**

| Parameters | Values |
|---|---|
| Frame Size | 1500Bytes |
| Time Slot | 0.2 msec |
| Channel Capacity | 10 Mbps |
| Learning Rate ($\alpha$) | 1.0 |
| History Length ($H_t$) | 15 |
| Discount Factor ($\gamma$) | 0.8, 0.9 |

**Table 3: Simulation scenarios**

| # | Nodes | Load | Noise |
|---|---|---|---|
| 1 | 5 | Low | No |
| 2 | 5 | Low | Yes |
| 3 | 15 | Average | No |
| 4 | 15 | Average | Yes |
| 5 | 25 | High | No |
| 6 | 25 | High | Yes |
| 7 | 50 | Saturated | No |
| 8 | 50 | Saturated | Yes |

**Table 4: Blocks and their associated algorithm/mechanism/parameter**

| Building Block | Algorithm/Parameter | Default |
|---|---|---|
| Backoff | BEB, EIED | BEB |
| ACK | No ACK, ACK | ACK |
| Fragmentation (Fr) | 200, 500, 1000 bytes | 1500 bytes |
| Aggregation (Ag) | 2000 bytes | 1500 bytes |
| RTS/CTS | Enabled/Disabled | N/A |
| CW | $CW_{min} = 0 - 1023$ | $CW_{min} = 15$ |
| Carrier Sense (CS) | Enabled/Disabled | N/A |
| Data Trans. Rate (DR) | 6/9/12/24/36/48/54 (Mbps) | 54 Mbps |

### 6.4 DeepMAC versus IEEE 802.11

IEEE 802.11 is one of the most popular wireless protocol that is based on CSMA/CA (Carrier Sense Multiple Access/Collision Avoidance) mechanism, which is the contention-based medium access control base for many of the current wireless protocols. In this section, we are comparing the performance of `DeepMAC` with IEEE 802.11 and consequently CSMA/CA mechanism.

IEEE 802.11 uses Binary Exponential Backoff (BEB) technique to randomize each node attempt of transmitting in order to reduce collisions. However, IEEE 802.11 random Backoff is decentralized and unable to efficiently handle collisions. Therefore, the network throughput degrades when the number of competing nodes increases. Giving collisions cannot be detect, IEEE 802.11 uses ACK mechanism to determine the successful reception of a packet. In addition, the RTS/CTS mechanism is introduced that can effectively ameliorate the hidden node problem. Although exchanging these control packets (i.e., ACK, RTS/CTS) is useful for successful packet transmission, they could introduce extra overhead for bandwidth utilization. In the following, we describe the throughput gains of `DeepMAC` versus IEEE 802.11 DCF (and CSMA/CA) in two different traffic loads: Low and High with different number of nodes and the assumption of no noise. Such assumption helps to simplify the effect of *avalanche rate* when multiple transmission rates are available.

**Low traffic load** In the first experiment, illustrated in Figure 7(a), the low load traffic corresponds to varying number of nodes from 1 to 15 that covers scenarios #1 and #3 in Table 3. In these scenarios, every 3 seconds a new node joins the network, and the simulation duration lasts for 45 seconds. As illustrated in Figure 7(a), IEEE 802.11 fails to fully utilize the channel bandwidth, while `DeepMAC` protocol effectively adapts to the network load changes by selecting the appropriate set of building blocks. Looking closely at the graph, when there is only one node, `DeepMAC` performs slightly better than IEEE 802.11 since it removes extra overheads (ACK and CS) from design. Overall, `DeepMAC` improves channel throughput by ∼ 6%. By looking closely to the selected blocks by `DeepMAC` in Table 5 when the number of nodes is 3, we observe that the "No ACK" mechanism is selected along with "Aggregation" that both can enhance the throughput by reducing extra control frame overhead. Interestingly, DRL agent learned that when the load of the network is low, it could eliminate control packets (e.g., ACK) to increase the throughput of the channel. These observations may look trivial for a human expert, but it makes it interesting when a DRL agent is able to learn such intuition on its own. In our implementation, when "No ACK" mechanism is selected we assume that the transmission reliability (e.g., ACK) is handled in the upper layers (e.g., TCP protocol in Transport layer). However, this raises the issue of cross-layer optimization since an unreliable MAC may impact TCP performance in which TCP assumes a packet loss is due to congestion, while it could be due to wireless link interference. We discuss the cross-layer design optimization as one of our future direction and discuss it later in Section 7.

**Table 6: Blocks selected by `DeepMAC` agent**

| # | DR | BEB | EIEB | CS | CW | No ACK | ACK | Fr | Ag | RTS/CTS |
|---|---|---|---|---|---|---|---|---|---|---|
| 1 | 54 | | | ■ | 31 | ■ | | 1500 | 2000 | |
| 2 | 24 | | | ■ | 31 | ■ | | 1500 | 1500 | |
| 3 | 54 | | | | 15 | ■ | | 1500 | 1500 | |
| 4 | 48 | ■ | | | 15 | | ■ | 1500 | 1500 | ■ |
| 5 | 54 | | | | 15 | ■ | | 1500 | 1500 | |
| 6 | 24 | ■ | | | 15 | | ■ | 1000 | 1500 | ■ |
| 7 | 36 | | | | 15 | | ■ | 500 | 1500 | |
| 8 | 24 | ■ | | | 15 | | ■ | 500 | 1500 | ■ |

Legend: ■ Selected BB / □ Not Selected or Default Value

**High traffic load** In the second experiment, we consider a high load traffic that corresponds to scenarios #5 and #7 in Table 3. At the beginning of the experiment 25 nodes are competing for the channel. At every 2 seconds a new node joins until the number of contending nodes reaches 50 as shown in Figure 7(b). The starting point of this figure is the continuation of Figure 7(a). As we observe, `DeepMAC` performs better than IEEE 802.11 even when the number of the nodes increases to 50 ($\sim$ 2% higher throughput) when the performance of both of the approaches degrades. By looking at Table 5, we observe that `DeepMAC` has selected EIED (Exponential Increase Exponential Decrease) over BEB. This can be because the slower reduction rate helps improving saturation throughput. Besides, control packets are also selected by the agent probably to avoid collisions and consequent retransmissions of large data packets.

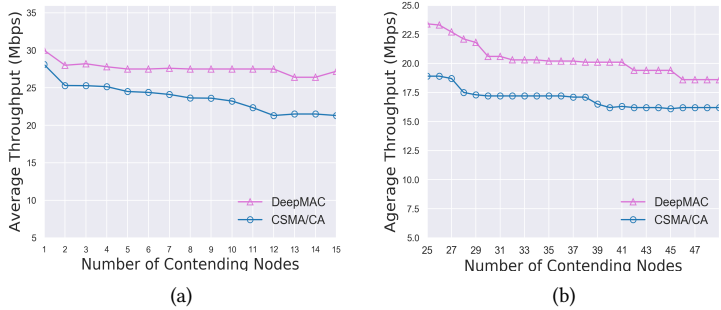

(a)  (b)

Figure 7: Throughput comparison of `DeepMAC` against CSMA/CA under (a) Low load traffic. (b) High load traffic.

Table 5: Selected Blocks by `DeepMAC` Under Different Networking Load

| # of Nodes | Traffic Load | Blocks Selected by DeepMAC |
|---|---|---|
| 3 | Low | No ACK, Aggregation |
| 15 | Average | ACK, Fragmentation, BEB, CW |
| 45 | High | EIED, CW, CS, ACK, Fragmentation, RTS/CTS |

## 6.5 Analysis of the Selected Blocks Under Different Scenarios

This subsection focuses on understanding the reasons behind the selected blocks by the agent under different scenarios. The selected blocks are shown in Table 6 in which the highlighted cells indicate the selected blocks in each scenario along with their selected values (If the block has a tunable parameter), while white cells indicate that the corresponding block is not selected in the given scenario. In the following, we divide our observations about `DeepMAC` behavior to three parts and discuss further about each case individually.

**Low load with/without noise** In scenarios with the low load when the noise is absent (e.g., Scenario #1). As shown in Table 6 cells corresponding to control packets such as ACK or RTS/CTS are not selected by the agent. This observation is justifiable. Even though the control packets are much smaller than the data packets, the time spent for control packet transmission is not negligible. Therefore, when the network is under saturated, and the number of competing nodes are small, the DRL agent avoids control packet overheads to maximize the throughput. Intuitively, to reduce the relative percentage of the time loss due to packet overhead and MAC coordination, frame aggregation is also selected by the agent. While for the same scenario, when the noise is present, it adds Career Sensing (CS) block. This may be due to the fact that the agent learns such a mechanism could be useful when the throughput drops.

**Average load with/ without noise** For scenarios with the average level of noise (Scenario #3 and #4) except common ACK mechanism selection, there is no obvious pattern. This observation could be either because such scenarios are not able to capture the useful information of what specific blocks should be selected, or it is simply because selecting different blocks does not provide a significant difference in the achieved throughput in such scenarios.

**High and saturated load with/without noise** We divide our observations for the three following scenarios: (1) The first observation in the high and saturated scenarios (Scenario #5 to #8) is the ACK mechanism that is selected by the agent. Intuitively, this could be because the agent learns such a mechanism can contribute to prevent more number of collisions and retransmissions to enhance the throughput. (2) When comparing scenario 5 to 6, we observe that the agent activates the Fragmentation block. The size of the sub-frames in practice plays an important factor that can influence network throughput performance for a given channel condition. The larger packets could contribute to the higher Packet Error Rate (PER) which would cause throughput drop due to a large number of retransmissions. (3) When the network is saturated, the agent selects protection mechanisms such as ACK and RTS/CTS along with smaller frame sizes and lower bitrate. However, it is not clearly obvious if the smaller frames contribute much to enhance the throughput since small fragments with the extra introduced overhead could also decrease the throughput performance.

The varying results reveal why it is extremely hard for an algorithm based on manually-specified rules and thresholds to capture the optimal solution, and why instead it is helpful to use machine learning techniques to optimize the design of control algorithms as well as, getting insights about what functionality (block) is useful under what scenario.

# 7 DISCUSSION/ FUTURE DIRECTIONS

The proposed protocol design framework opens up the possibility of future research directions in different dimensions. We discuss some of the directions in the following.

**Build a library of protocol elements** One of our main future directions is to develop and build an open library/database of protocol design building blocks to be utilized by us and the networking research community. We aim to target to identify and develop the basic building blocks of popular variants IEEE 802.11 standards (i.e., a/b/g/n/ac/ax).

**Get new insights about protocol design** One of the most attractive aspect of Machine Learning (ML) techniques lies in their ability to discover new knowledge that lies beyond the current human understanding and perception. Current protocol design approaches are limited to human perception and understanding of this field, thus limiting the potential for extracting new and unexpected insights during the design process. Therefore, by applying ML techniques to protocol design, one may clearly see an exceptional potential in developing more efficient, robust, and autonomous systems for protocol design. Additionally, such a combination of two fields could lead to new understanding of ML techniques applied to challenging and evolving data, as well as gaining invaluable insights into how ML may advance the future of network communication field.

**Deal with heterogeneity** Today, billions of wireless devices each of which has its own specification and characteristics are competing for spectrum, creating complex systems in which existing human-driven network design strategies are performance-wise inefficient. There are some recent approaches towards managing such heterogeneous networking environments (e.g., DARPA Spectrum Collaboration Challenge (SC2) [1]), but these visions focus only on a narrow spectrum. We envision a framework that not only considers the MAC and physical layer, but the network protocol stack as a whole, allowing devices to collectively learn from each other and transfer knowledge among each other. This will form intelligent and adaptive system for protocol design that is capable of self-management, even when dealing with a plethora of diverse devices.

**Cross-layer design optimization** In our implementation, we assume that the transmission reliability (e.g., ACK) is handled in the upper layers (e.g., TCP protocol in Transport layer). An unreliable MAC protocol may impact the performance of upper-layer protocols ( e.g., TCP as it assumes a packet loss is due to the congestion while it could be caused by wireless link poor quality). Therefore, it would be more efficient to have a cross-layer design optimization where constraints from different layers are enforced to select certain blocks to optimize the whole network stack.

**Make protocols more robust** How well can the learned policy perform in conditions never seen during training? This is a question about the generalization capability of the learning algorithm [54]. The huge diversity of possible network conditions implies that even a protocol that works well across a wide variety of network conditions may suffer from bad performance on other networks. Thus, enhancing the robustness of protocols is clearly desirable. This typically involves identifying scenarios that result in poor performance by the protocol and using these scenarios for debugging and for guiding changes to the protocol. The search for challenging network conditions must be customized for the specific protocol under consideration. Bad protocol behavior might be triggered by complex sequences of changes in network conditions, making identifying such examples challenging. Network conditions that have been shown to induce bad performance for a protocol will also contain hints regarding where the problem lies, i.e., the demonstrated problem should be explainable. One possible way to identify bad protocol performance is to leverage RL to generate adversarial network traces for an input protocol by observing protocol behavior and adaptively changing its network conditions to harm its performance relative to the optimal. Then, the adversarial network traces generated by the adversarial framework can be leveraged to train more robust RL-based protocols.

The ultimate goal of this study is to train network nodes to build their own customized new MAC protocols. Recent research [30, 41] suggests this is a hard goal to achieve when all agents (i.e., nodes) start with no previous knowledge. Consequently, scheduled training that alternates between supervised learning and self-play as suggested in [42] seems promising to emerge fully new protocols.

# 8 CONCLUSION

In this paper, we motivated the importance of a shift from the human-driven protocol design process to a machine-based design. We proposed and evaluated a framework for MAC protocol design optimization using a DRL-based approach. We have shown that by observing the decisions of the `DeepMAC` agent and using a method such as input modularization (protocol decomposition into building blocks), it is possible to extract information about the associated block selection by the agent. We envision this method could offer useful insights, especially to protocol designers to build deeper perception about the significance of an individual or a set of protocol blocks (functions) under different scenarios. This could help them focusing on enhancements/modifications of important components than focusing on the whole protocol performance in order to enhance the protocol design and performance.